\documentclass[11pt,oneside]{article}
\usepackage[margin=1in]{geometry}
\usepackage[T1]{fontenc}
\usepackage[utf8]{inputenc}
\usepackage{relsize}

\usepackage{amssymb}
\usepackage{amsmath}
\usepackage{graphicx}
\usepackage{url}
\usepackage{xspace}
\usepackage{epstopdf}
\usepackage[leftcaption]{sidecap}
\sidecaptionvpos{figure}{c} 

\usepackage{color}
\let\hat\widehat
\let\tilde\widetilde

\newcommand{\E}{\mbox{$\mathbb{E}$}}

\title{\textbf{PHYSTAT Informal Review:\\
Marginalizing versus Profiling of Nuisance Parameters}\\
\vspace{0.5cm}
\smaller{}Notes by\\
\vspace{0.5cm}
Robert D. Cousins\\
Department of Physics and Astronomy\\
University of California, Los Angeles 90095\\
Los Angeles, CA\\
\vspace{0.5cm}
and \\
\vspace{0.5cm}
Larry Wasserman\\
Department of Statistics and Data Science\\
Carnegie Mellon University\\
Pittsburgh, PA 15213}

\usepackage[sectionbib]{natbib}
\bibpunct{(}{)}{;}{a}{,}{,}  
\usepackage{chapterbib}

\begin{document}

\maketitle

\begin{abstract}
This is a writeup, with some elaboration, of the talks by the two authors (a physicist and a statistician) at the first PHYSTAT Informal review on January 24, 2024. We discuss Bayesian and frequentist approaches to dealing with nuisance parameters, in particular, integrated versus profiled likelihood methods. In regular models, with finitely many parameters and large sample sizes, the two approaches are asymptotically equivalent.  But, outside this setting, the two methods can lead to different tests and confidence intervals.  Assessing which approach is better generally requires comparing the power of the tests or the length of the confidence intervals. This analysis has to be conducted on a case-by-case basis. In the extreme case where the number of nuisance parameters is very large, possibly infinite, neither approach may be useful. Part I provides an informal history of usage in high energy particle physics, including a simple illustrative example.  Part II includes an overview of some more recently developed methods in the statistics literature, including methods applicable when the use of the likelihood function is problematic.
\end{abstract}

\thispagestyle{empty} 

\newpage
\pagenumbering{arabic} 
\section*{\centering Foreword}
This document collects reviews written by Bob Cousins, Distinguished Professor Emeritus in the Department of Physics and Astronomy at UCLA, and Larry Wasserman, Professor of Statistics and Data Science at Carnegie Mellon University, on inferential methods for handling nuisance parameters.

These notes expand upon the main points of discussion that arose during the first of a new event series called PHYSTAT–Informal Reviews, which aims to promote dialogue between physicists and statisticians on specific topics. The inaugural event held virtually on January 24, 2024, focused on the topic of "Hybrid Frequentist and Bayesian approaches". Bob Cousins and Larry Wasserman presented excellent talks, and the meeting was further enriched by a valuable discussion involving the statisticians and physicists in the audience.

In Part I, Bob Cousins provides a comprehensive overview of how high-energy physicists have historically tackled the problem of inferring a parameter of interest in the presence of nuisance parameters. The note offers insights into why methods like profile likelihood are somewhat dominant in practice while others have not been widely adopted. Various statistical solutions are also compared through a classical example with relevance in both physics and astronomy. This memorandum is a wealth of crucial references concerning statistical practice in the physical sciences.

In Part II, Larry Wasserman examines the problem of conducting statistical inference with nuisance parameters from the perspectives of both classical and modern statistics. He notes that standard inferential solutions require a likelihood function, regularity conditions, and large samples. Nonetheless, recent statistical developments may be used to ensure correct coverage while relaxing these requirements. This review offers an accessible self-contained overview of some of these approaches and guides the reader through their strengths and limitations.

Together, the written accounts below provide a cohesive review of the tools currently adopted in the practice of statistical inference in searches for new physics, as well as their historical and technical justification. They also offer a point of reflection on the role that more sophisticated statistical methods and recent discoveries could play in assisting the physics community in tackling both classical and new statistical challenges.

The organizing committee of PHYSTAT activities is grateful to both authors for their engagement and valuable contributions, which we believe will benefit both the statistics and physics communities.

\vspace{0.5cm}
\hfill\begin{raggedleft}
\emph{Sara Algeri,}
\end{raggedleft}

\hfill\begin{raggedleft}
School of Statistics, University of Minnesota, Minneapolis, MN
\end{raggedleft}

\vspace{0.1cm}
\hfill\begin{raggedleft}On behalf of the PHYSTAT Organizing Committee
\end{raggedleft}

\newpage
\begin{center}
{\bf \LARGE Part I: Physicist’s View}\\
\vspace{0.2cm}
by Robert D. Cousins\\

\end{center}

\newcommand{\hzero}{\ensuremath{\mathrm{H}_0}\xspace}

\newcommand\lhood{{\cal L}}
\newcommand\non{n_\textnormal{\scriptsize on}}       
\newcommand\ntot{n_\textnormal{\scriptsize tot}}     
\newcommand\binp{\rho}            

\newcommand\noff{n_{\rm off}}     
\newcommand\mus{\mu_{\rm s}} 
\newcommand\mub{\mu_{\rm b}}  
\newcommand\hathatmub{\hat{\hat{\mu}}_{\rm b}}
\newcommand\muon{\mu_{\rm on}}    
\newcommand\muoff{\mu_{\rm off}}  %
\newcommand\mutot{\mu_{\rm tot}}  

\newcommand\pbi{p_{\rm Bi}}
\newcommand\zbi{Z_{\rm Bi}}

\newcommand{\pP}{\ensuremath{p_\mathrm{P}}\xspace}
\newcommand{\zP}{Z_{\rm P}}

\newcommand{\pbf}{\ensuremath{p_\mathrm{BF}}\xspace}
\newcommand{\zbf}{\ensuremath{Z_\mathrm{BF}}\xspace}

\newcommand{\ppb}{\ensuremath{p_\mathrm{PB}}\xspace}
\newcommand{\zpb}{\ensuremath{Z_\mathrm{PB}}\xspace}

\newcommand\pgamma{p_\Gamma}
\newcommand\zgamma{Z_\Gamma}

\newcommand{\ppl}{\ensuremath{p_\mathrm{PL}}\xspace}
\newcommand{\zpl}{\ensuremath{Z_\mathrm{PL}}\xspace}

\newcommand{\zmidp}{\ensuremath{Z_\mathrm{mid\text{-}P}}\xspace}

\newcommand\zthresh{Z_{\rm thresh}}
\newcommand\zclaim{Z_{\rm claim}}
\newcommand\ztrue{Z_{\rm true}}
\newcommand\zdiff{\ztrue-\zclaim}

\section{Introduction}
This is a writeup (with some changes) of the 20-minute talk that I gave at a virtual PhyStat meeting that featured ``mini-reviews'' of the topic by a physicist \citep{cousinsphystat2023} and a statistician~\citep{wassermanphystat2023} Mine was largely based on my talk the previous year at a BIRS workshop at Banff~\citep{cousinsbanff2023}. This is a short personal perspective through the lens of sometimes-hazy recollections of developments during the last 40 years in high energy physics, and is not meant to be a definitive history. I also refer at times to my lecture notes from the 2018 Hadron Collider Physics Summer School at Fermilab~\citep{cousinshcpss2018}, which provide a more detailed introduction to some of the fundamentals, such as the duality of frequentist hypothesis testing and confidence intervals (``inverting a test" to obtain intervals).

Parametric statistical inference requires a statistical model, $p(y;\theta)$, which gives the probability (density) for observing data $y$, given parameter(s) $\theta$. In a given context, $\theta$ is partitioned as $(\mu,\beta)$, where $\mu$ is the parameter of interest (taken to be a scalar in this note, e.g., the magnitude of a putative signal), and $\beta$ is typically a vector of ``nuisance parameters'' with values constrained by subsidiary measurements, usually referred to as ``systematic uncertainties'' in particle physics (e.g., detector calibration ``constants'', magnitudes of background processes, etc.). 

This note discusses the common case in which one desires an algorithm for obtaining ``frequentist confidence intervals (CI)'' for $\mu$ that are valid at a specified confidence level (CL), no matter what are the unknown values of $\beta$. Ideally, CIs have exact ``frequentist coverage'' of the unknown true value of $\mu$, namely that in imagined repeated applications of the algorithm for different data $y$ sampled from the model, in the long run a fraction CL of the confidence intervals will contain (cover) the unknown true $\mu$. In practice, this is often only approximately true, and some over-coverage (fraction higher than CL) is typically tolerated (with attendant loss of power), while material under-coverage is considered disqualifying. 

The construction of confidence intervals requires that for every possible true $\mu$, one chooses an ordering of the sample space $y$. This is done by specifying an ordered ``test statistic'', a function of the observed data and parameters. In the absence of nuisance parameters, standard choices have existed since the 1930s. Thus, a key question to be discussed is how to treat (``eliminate") the nuisance parameters in the test statistic, so that the problem is reduced to the case without nuisance parameters.

A second question to be discussed is how to obtain the sampling distribution of the test statistic under the statistical model(s) being considered. There is a vast literature on approximate analytic methods that become more accurate as the sample size increases.  For small sample sizes in particle physics, Monte Carlo simulation is typically used to obtain these distributions.  The crucial question is then, which values(s) of the nuisance parameters should be used in the simulations, so that the inferred confidence intervals for the parameter of interest have valid coverage, no matter what are the true values of the nuisance parameters.

Section~\ref{teststat} defines the profile and marginalized likelihoods.
Section~\ref{sampling} introduces the two corresponding ways to treat nuisance parameters in simulation, the parametric bootstrap and marginalization (sampling from the posterior distribution of the nuisance parameters).
Section~\ref{example} is a longer section going into detail regarding five ways to calculate the significance $Z$ in the on/off problem (ratio of Poisson means).
Section~\ref{sec:davison} briefly highlights the talk by Anthony Davison on this subject at the recent 2023 BIRS workshop at Banff.
Section~\ref{final} contains some final remarks.
\section{Choice of test statistic}
\label{teststat}

The observed data $y$ are plugged into the statistical model $p(y;\theta)$ to obtain the likelihood function $\lhood(\mu,\beta)$. The global maximum is at 
$\lhood(\hat\mu,\hat\beta)$. The ``profile likelihood function" (of {\em only} $\mu$) is
\begin{equation}
\lhood(\mu,\hat{\hat{\beta}}),
\end{equation}
where, for a given $\mu$, the restricted estimate $\hat{\hat\beta}$ maximizes $\lhood$ for that $\mu$. It is also written as
\begin{equation}
  \sup_\beta \lhood(\mu,\beta).
\end{equation} (The restricted estimate is denoted by $\hat\beta_\mu$ in~\cite{reidslac2003}.)  
It is common in HEP to use as a test statistic the ``profile likelihood ratio'' (PLR)  with the likelihood at the global maximum in the denominator,
\begin{equation}
\frac{\lhood(\mu,\hat{\hat\beta})}{\lhood(\hat\mu,\hat\beta)}.
\end{equation}

The ``integrated'' (or ``marginalized'') likelihood function (also of only $\mu$) is
\begin{equation}
\int \lhood(\mu,\beta) \pi(\beta)  d\beta,
\end{equation}
where $\pi(\beta)$ is a weight function in the spirit of a Bayesian prior pdf. (More generally, it could be $\pi(\beta|\mu)$.)

One can then use either of these as if it were a 1D likelihood function $\lhood(\mu)$ with the nuisance parameters ``eliminated''.  The question remains, what are the frequentist properties of intervals constructed for $\mu$? The long tradition in particle physics is to use the profile likelihood function and the asymptotic theorem of \cite{wilks} to obtain approximate 1D confidence intervals (as well as multi-dimensional confidence regions in cases beyond the scope of this note). For a brief review, see Section 40.3.2.1 and Table 40.2 of the~\cite{pdg2022}. For decades, a commonly used software tool has been the classic program (originally in FORTRAN and later in C and C++), “Minuit: A System for Function Minimization and Analysis of the Parameter Errors and Correlations''~\citep{minuit}.
MINUIT refers to the uncertainties from profile likelihood ratios as “MINOS errors”, which are discussed by \cite{minoscpc}. I believe that the name “profile likelihood” entered the HEP mainstream in 2000, thanks to a talk by statistician Wolfgang Rolke on work with Angel L\'opez at the second meeting of what became the PhyStat series~\citep{rolkeclk,rolkelopez2001,Rolke2005}. Interestingly,~\cite{minoscpc} and \cite{rolkeclk} conceptualize the same math in two different ways, as discussed in my lectures~\citep{cousinshcpss2018}; this may have delayed recognizing the connection (which was not immediate, even after Wolfgang's talk).

At some point, a few people started integrating out nuisance parameter(s), typically when there was a Gaussian/normal contribution to the likelihood, while treating the parameter of interest in a frequentist manner.  One of the earlier examples (citing yet earlier examples) was by myself and Virgil Highland, “Incorporating systematic uncertainties into an upper limit”~\citep{cousinshighland}.
We looked at the case of a physics quantity (cross section) $\sigma = \mub / L$,
where $\mub$ is an unknown Poisson mean and $L$ is a factor called the integrated luminosity. One observes a sampled value $n$ from the Poisson distribution. 
From $n$, one can obtain a usual frequentist upper confidence limit on $\mub$. 
If $L$ is known exactly, one then simply scales this by $1/L$ to get an upper confidence limit on $\sigma$. We considered the case in which, instead, one has an independent unbiased measurement of $L$ with some uncertainty, yielding a likelihood function of $\lhood(L)$. Multiplying $\lhood(L)$ by a prior pdf for $L$ yields a Bayesian posterior pdf for $L$. We advocated using this as a weight function for what we now call integrating/marginalizing over the nuisance parameter $L$. A power series expansion gave useful approximations.

Our motivation was to ameliorate an effect whereby confidence intervals derived from a discrete observable become shorter when a continuous nuisance parameter is added to the model.  In the initially submitted draft, we did not understand that we were effectively grafting a Bayesian pdf onto a frequentist Poisson upper limit.  Fred James sorted us out before publication. A more enlightened discussion is in my paper with Jim Linnemann and Jordan Tucker~(\citeyear{CLT}), referred to as CLT, discussed below. Luc Demortier noted that, in these simple cases, what we did was the same math as Bayesian statistician George Box’s prior predictive $p$-value~\citep{box1980,CLT}.  Box calculated a tail probability after obtaining a Bayesian pdf, and we averaged a frequentist tail probability over a Bayesian pdf. The two calculations simply reverse the order of two integrals.

A key paper was submitted in the first year of LHC data taking, by
Glen Cowan, Kyle Cranmer, Eilam Gross, and Ofer Vitells (CCGV), “Asymptotic formulae for likelihood-based tests of new physics”~(\citeyear{ccgv2011}).  
They provided useful asymptotic formulas for the PLR and their preferred variants, which have been widely used at the LHC and beyond. I believe that this had a significant side effect: since asymptotic distributions were not known in HEP for other test statistics, the default became PLR variants in CCGV, even for small sample size, in order to have consistency between small and large sample sizes.

Meanwhile, in the statistics literature, there is a long history of criticism of the simple PLR. At PhyStat meetings, various higher-order approximations have been discussed by Nancy Reid at SLAC in~\citeyear{reidslac2003}; by Nancy~\citep{reidoxford2005,reidoxford2} and me~\citep{cousinsoxford2005} at Oxford in 2005; by Anthony Davidson in the Banff Challenge of 2006~\citep{davisonbanffchallenge}, at PhyStat-nu 2019~(\citeauthor{davison2019}), and at Banff last year~\cite{davisonbanff2023}; and
by statistician Alessandra Brazzale~(\citeyear{brazzalephystat2019}) and physicist Igor Volobouev and statistician Alex Trindale at PhyStat-DM in~\citeyear{volobouevphystat2019}. Virtually none of these developments have been adopted for widespread use in HEP, as far as I know.

Bayesian statisticians at PhyStat meetings have long advocated marginalizing nuisance parameters, even if we stick to frequentist treatment of the parameter of interest, e.g., James Berger, Brunero Liseo, and Robert Wolpert, “Integrated Likelihood Methods for Eliminating Nuisance Parameters”, (\citeyear{berger1999}). 
Berger et al., in their rejoinder to a comment by Edward Susko, wrote,
“Dr. Susko ﬁnishes by suggesting that it might be useful to compute both proﬁle and integrated likelihood answers in application, as a type of sensitivity study. It is probably the case that, if the two answers agree, then one can feel relatively assured in the validity of the answer. It is less clear what to think in the case of disagreement, however. If the answers are relatively precise and quite different, we would simply suspect that it is a situation with a ‘bad’ proﬁle likelihood.” I agree with Susko’s suggestion to compute (and report) both proﬁle and integrated likelihood answers, to build up experience and to see if Berger et al. are right in real cases relevant to HEP. 

However, in my experience, marginalizing nuisance parameters is relatively rare at the LHC.  This is for various reasons, in my opinion, including:
\begin{itemize}
\item The historical traditions of frequentist statistics in HEP, and the naturalness of PLRs as test statistics;
\item  In particular, the  historical precedent of MINUIT MINOS;
\item The (false) impression that doing something Bayesian-inspired is unjustifiable to a frequentist, when frequentist calibration measures can provide the justification.
\item The asymptotic formulas of CCGV are readily available for profile likelihood but not for marginalization, and are the default in the now-dominant software tools.
\end{itemize}
It also seems common that the minimization required for profiling is less CPU-intensive than the integration required for marginalization.

Thus, the profile likelihood ratio (including variants with parameters near boundaries as in CCGV~\citep{ccgv2011}) is the most common test statistic in use today at the LHC, and perhaps in all of HEP. Among other uses, it is the basis of the ordering for confidence intervals advocated by Gary Feldman and myself~(\citeyear{feldman1998}), which turned out to be the no-nuisance-parameter special case of inversion of the “exact” PLR test in the treatise of “Kendall and Stuart” and later versions~\citep{kendall1999}. I recall a long discussion between Gary Feldman and Kyle Cranmer at a pub at the Oxford Phystat in 2005 regarding how to interpret Eqs. 22.4-22.7 of \cite{kendall1999}.  For part of that historical context, see  Kyle’s PhyStat contributions in \citeyear{cranmerslac2003} (Sec.~9) and \citeyear{cranmeroxford2005} (Sec.~4.3), as well as the contribution by Giovanni Punzi in \citeyear{punzioxford2005}.

Various studies have been done regarding the elimination of nuisance parameters
in the test statistic (typically a likelihood ratio), many concluding
that results are relatively insensitive to profiling vs
marginalization, so that the choice can be made based on CPU time.  See,
for example, John Conway's talk and writeup at
PhyStat in \citeyear{conwayphystat2011}.
Jan Conrad and Fredrik Tegenfeldt studied cases with favorable coverage~\citep{conradtegenfeldt2005,conradphystat2005}. Notably, as Kyle Cranmer showed at Oxford in \citeyear{cranmeroxford2005}, for Gaussian uncertainty on the mean $\mub$, marginalizing the nuisance parameter can be badly anti-conservative (have severe undercoverage) when calculating signal significance (a different situation than the upper limits considered by myself and Highland (\citeyear{cousinshighland})). This was one of the motivations of the studies by CLT.  

\section{Sampling distributions of the test statistic(s) under various \\ hypotheses}
\label{sampling}

Rather than pursuing higher-order corrections to the PLR, the general practice in HEP has been to stick with the PLR test statistic, and to
use Monte Carlo simulation (known as “toy MC”) to obtain the finite-sample-size distribution(s) of the PLR under the null hypothesis, and under alternative(s) as desired. These results are often compared to the relevant asymptotic formulas, typically from CCGV.

So now one has the question: {\em How should nuisance parameters be treated
in the simulations?}  We need to keep in mind that we want correct frequentist coverage
of the parameter of interest when nature is sampling the data using the {\em unknown
true values} of the nuisance parameters.  An important constraint (at
least until now) is that it is generally thought to be impractical to perform
MC simulation for each of many different sets of values of the nuisance
parameters (especially in high dimensions).  So how should one judiciously choose those values to be used in simulation?

The usual procedure (partially based on a desire to be “fully frequentist”), is to “throw toys” using the profiled values of the nuisance parameters, i.e., their ML estimates conditional on whatever value of the parameter(s) of interest are being tested.
At some point, this procedure was identified as the parametric bootstrap. (The first time that I recall was in an email from Luc Demortier in 2011, pointing to his contribution to the PhyStat-LHC 2007 Proceedings \citep{luc2007phystat} and to the book by Bradley Efron and Robert Tibshirani (\citeyear{efrontibshirani}).) The hope is that the profiled values of the nuisance parameters are a good enough proxy for the unknown true values, even for small sample size. I have since learned that there is a large literature on the parametric bootstrap and its convergence properties. See, for example, Luc's talk \citep{luc2012slac}, which describes a number of variants, with references, as well as Anthony Davison's talk at Banff~\citeyear{davisonbanff2023}, emphasized in Section~\ref{sec:davison}.

In an alternative procedure, sometimes called the Bayesian-Frequentist hybrid, in each pseudo-experiment in the toy-throwing, a new set of values of the unknown nuisance parameters is sampled from their posterior distributions, and used for sampling the data. The parameter of interest, however, continues to be treated in a frequentist fashion (confidence intervals, $p$-values, etc.). More details are in Section~\ref{BF} below.

I would like to see more comparisons of these results with alternatives, particularly those from marginalizing the nuisance parameters with some judiciously chosen priors. I find it a bit comforting to take an average over a set of nuisance parameters in the neighborhood of the profiled values. Coverage studies can give guidance on the weight function $\pi(\beta)$ used in the marginalization, presumably considering default priors thought to lead to good coverage.

\section{Example: ratio of Poisson means}
\label{example}

I “cherry-picked” an example that appears often in the statistical literature; 
is an important prototype in HEP and astronomy; and is the topic of a paper for which I happen to be a co-author, including an interesting result for marginalization. This is the ratio of Poisson means, which maps onto the “on/off” problem in astronomy and the “signal region plus sideband” problem in HEP. I focus here mostly on the concepts, as the algebra is in the CLT paper~\citep{CLT}, much of which was inspired by Jim Linnemann's talk at the SLAC PhyStat in \citeyear{linnemannslac2003}.

\begin{SCfigure}[2.95][htbp]
\label{onoff}
\caption{The two bins of the on/off problem, with observed contents $\non$ and $\noff$. The “on” bin has (potentially) signal and background events with unknown Poisson means $\mus$ and $\mub$, respectively, with total Poisson mean $\mus + \mub$. The “off” bin has only background with unknown Poisson mean $\muoff$.  In this simplest example, the ratio $\tau = \muoff / \mub$  is assumed known.}
\includegraphics*[width=1.5in]{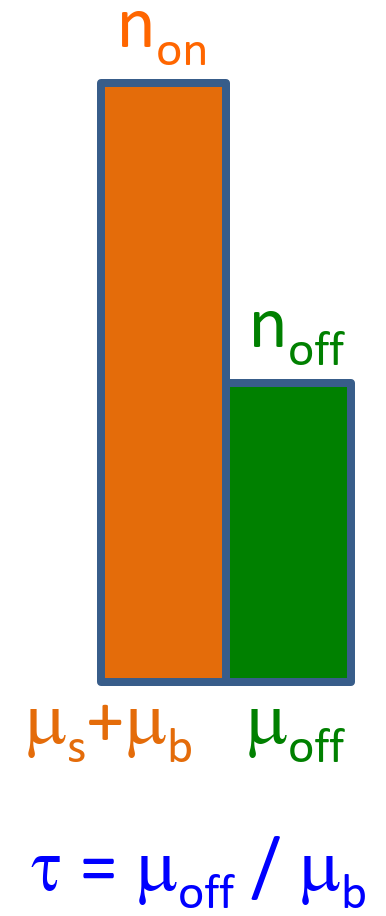}
\end{SCfigure}

As in Fig.~\ref{onoff}, we have a histogram with two bins; in astronomy, these are counts with the telescope ``on'' and ``off'' the putative source, with observed contents $\non$ and $\noff$, respectively. In HEP, there is a bin in which one looks for a signal, and a ``sideband'' bin that is presumed to have background but no signal.  So the “on” bin has (potentially) signal and background events with unknown Poisson means $\mus$ and $\mub$, respectively, with total Poisson mean $\mus + \mub$. The “off” bin has only background with unknown Poisson mean $\muoff$. The ratio of the total Poisson means in the two bins is denoted by $\lambda$,
\begin{equation}
  \lambda = \frac{\muoff}{\mus+\mub}.
\end{equation}

In this simple version discussed here, we assume that the ratio of the Poisson means of background in the two bins, $\tau = \muoff / \mub$,  is precisely known. In astronomy, this could be the ratio of observing times; in HEP, one might rely on simulation or control regions in the data. (If $\tau$ is determined with non-negligible uncertainty from two more control regions, as in the so-called ABCD method in HEP, more issues arise.)
In a search for a signal, we test the null hypothesis \hzero: $\mus = 0$, 
which we can rephrase as \hzero: $\lambda = \tau$. One can choose the nuisance parameter to be either $\muoff$ or $\mub$, and we choose the latter.

\subsection{\boldmath $\zbi$: conditioning on $\ntot$ and using Clopper-Pearson intervals}
The standard frequentist solution, invoking ``conditioning'' \citep{jamesroos,reid1995,CLT}, is based on the fact that the total number of counts, $\ntot = \non + \noff$, provides information about the uncertainty on the ratio of Poisson means, but no information about the ratio itself.  Thus, $\ntot$ is known as an ancillary statistic, and most statisticians agree that one should treat $\ntot$ as fixed, i.e., calculate (binomial) probabilities of $\non$ and $\noff$ conditional on the observed $\ntot$. We can again rephrase the null hypothesis in terms of the binomial parameter $\rho$, namely \hzero: $\rho  = \mub / (\mub + \tau \mub) = 1 / (1+\tau)$. So we do a binomial test of $\rho = 1 / (1+\tau)$, given data $\non$, $\ntot$. The $p$-value $\pbi$ is obtained by inversion of binomial confidence intervals.

There are many available choices for sets of binomial confidence intervals. For $\pbi$, CLT adheres to the standard in HEP and follows the ``exact'' construction of \cite{clopper1934}, with the guarantee of no undercoverage. Note, however, that ``exact'', as commonly used in situations such as this (going back to Fisher), means that a small-sample construction (as opposed to asymptotic theory) was used, but beware that in a discrete situation such as ours, there will be overcoverage, sometimes severe.

The $p$-value $\pbi$ is converted to a $Z$-value (sometimes called $S$ in HEP) with a 1-tailed Gaussian convention to obtain $\zbi$. (As in CLT, $Z = \sqrt{2}\, \text{erf}^{-1}(1-2p)$.

\subsection{\boldmath $\zmidp$: conditioning on $\ntot$ and using Lancaster mid-P intervals}

After CLT, a paper by myself, Hymes, and Tucker (\citeyear{cht2010}) advocated using Lancaster's mid-P binomial confidence intervals~\citep{lancaster61} to obtain confidence intervals for the ratio of Poisson means. For any fixed $\ntot$, the mid-P intervals have over-coverage for some values of the binomial parameter and undercoverage for other values.  In the {\em unconditional} ensemble that includes $\ntot$ sampled from the Poisson distribution with the true total mean, the randomness in $\ntot$ leads to excellent coverage of the true ratio of Poisson means.

\subsection{\boldmath $\zpl$: profile likelihood and asymptotic theory}
The asymptotic result from Wilks’s Theorem (details in Section 5 of CLT and in \cite{pdg2022}, Section 5) yields the $Z$-value $\zpl$ from the profile likelihood function. 

\subsection{\boldmath $\zbf$: marginalization of nuisance parameter $\mub$ \\(Bayesian-frequentist hybrid)}
\label{BF}

First, consider the case when $\mub$ is known exactly. The $p$-value (denoted by $p_{\rm P}$) for \hzero: $\mus = 0$ is the Poisson probability of obtaining $\non$ or more events in the “on” bin with true mean $\mub$.

Then, introduce uncertainty in $\mub$ in a Bayesian-like way, with uniform prior for $\mub$ and likelihood $\lhood(\mub)$ from the Poisson probability of observing $\noff$ in the “off” bin, thus obtaining posterior probability $P(\mub|\noff)$, a Gamma distribution.
Finally, take the weighted average of \pP over $P(\mub|\noff)$ to obtain the Bayesian-frequentist hybrid $p$-value \pbf:
\begin{equation}
  \pbf = \int p_{\rm P} P(\mub|\noff) d\mub,
\label{pbfint}
\end{equation}
and map to $\zbf$ (called $\zgamma$ by CLT).

Jim Linnemann discovered numerically, and then proved, that $\zbf = \zbi$ (!). His proof is in Appendix C of CLT. This justifies, from a frequentist point of view, the use of the uniform prior for $\mub$, which is on shaky ground in real Bayesian theory.

The integral in Eq.~\ref{pbfint} can be performed using the Monte Carlo method as follows.
One samples $\mub$ from the posterior pdf $P(\mub|\noff)$, and computes the frequentist $p_{\rm P}$ for each sampled value of $\mub$ (using same observed $\non$). 
Then one calculates the arithmetic mean of these values of $\pP$ to obtain $\pbf$, and converts to $\zbf$ as desired.  (Note that $\pP$’s are averaged, not $\zP$’s.)

\subsection{\boldmath $\zpb$: parametric bootstrap for nuisance parameter $\mub$}

For testing \hzero: $\mus = 0$, we first find the profile likelihood estimate of $\mub$. See \cite{lima}, cited by CLT, for the mathematics.  

Conceptually, we use the data in both bins by noting as above that for $\mus=0$,
$\ntot = \non + \noff$ is a sample from Poisson mean $\mub+\muoff = \mub(1 + \tau)$.
Solving for $\mub$, the profiled MLE of $\mub$ for $\mus=0$ is $\hathatmub =(\non + \noff ) / (1 + \tau)$. The parametric bootstrap takes this value of $\mub$ as truth and proceeds to calculate the $p$-value $\ppb$ as in $p_{\rm P}$ above.  This can be done by simulation or direct calculation, leading to $\zpb$.

\subsection{Numerical examples}

The calculated $Z$-values for the various recipes are shown in Table 1 for three chosen sets of values of $\tau$, $\non$, and $\noff$.

Explaining the first set in detail, suppose $\tau = 1.0$, $\non  = 10$, $\noff = 2$.
Following the ROOT commands in Appendix E of CLT yields the “Exact” $\zbi = \zbf = 2.07$.
Alternatively, one can use the ROOT \citep{tefficiency} class TEfficiency for two-tailed binomial confidence intervals. Recall that we need a one-tailed binomial test of $\rho = 1 / (1+\tau)$, given data $\non$, $\ntot$.  Applying the duality between tests and intervals~\citep{cousinshcpss2018}, we seek the confidence level (CL) of the confidence interval (CI) that just includes the value of $\rho = 1 / (1+1) = 0.5$.  By iterating $p$ and the $\mathrm{CL}=1-2*p$ in TEfficiency manually until obtaining the endpoint=0.5 (for $\mathrm{CL} = p=0.019287$), I arrived at the following ROOT commands (converting $p=0.019287$ to $Z$ as in CLT). 
\begin{verbatim}
double n_on=10
double n_off=2
double p=0.019287
endpoint = TEfficiency::ClopperPearson(n_on+n_off,n_on,1.-2.*p,false);
endpoint
\end{verbatim}
Then I could switch to TEfficiency::MidPInterval to get the Lancaster mid-P interval and similarly obtain $\zmidp = 2.28$.

The asymptotic result from Wilks’s Theorem gives the value from the profile likelihood function, $\zpl = 2.41$.

For the parametric bootstrap, for $\mus = 0$ and $\ntot=12$, the profiled MLE is $\hathatmub = 6$, and so $\muoff = 6$. So I generated toys with $\mu = 6$ in each bin (!), to find  $\zpb = 2.32$.

The results for the second set of values ($\tau = 2.0$, $\non  = 10$, and $\noff = 2$) are found similarly. 
For the parametric bootstrap, for $\mus = 0$, the profiled MLE is $\hathatmub = 4$, and so $\muoff = 8$.
So I generated toys with $\mu = 4, 8$ in the bins where we observed 10, 2 (!).  $\zpb = 3.46$.

The third set of values (added since my talk) has no events in the off sideband, $\noff=0$. In the discussion following our talks, I expressed concern about generating toys for the parametric bootstrap in this situation.  In fact, it is not a problem.  With $\tau = 2.0$ and $\non  = 6$, then for $\mus = 0$ and $\ntot=6$, the profiled MLE is $\hathatmub = 2$, and so $\muoff = 4$.
So I generated toys with $\mu = 2, 4$ in the bins where we observed 6, 0 (!).  $\zpb = 3.41$ (In more complicated models, zero events observed in multiple control regions might be an issue.)

\begin{table}
   \renewcommand*{\arraystretch}{1.2}
  \centering
  \label{tab:z}
  \begin{tabular}{lccc}
  \hline  
    $\tau$                          &   1.0       & 2.0     &  2.0    \\ 
    $\non$                          &   10      & 10    &  6    \\
    $\noff$                         &   2       & 2     &  0    \\
    $\ntot$                         &   12      & 12    &  6    \\
    $\hathatmub$ for $\mus=0$ &    6      & 4     &  2    \\
    $\zbi=\zbf$, Clopper-Pearson    &   2.07    & 3.27  &  3.00  \\
    $\zmidp$, Lancaster mid-P         &   2.28    & 3.44 &   3.20 \\
    $\zpl$                          &   2.41    & 3.58 & 3.63   \\ 
    $\zpb$                          &   2.32    & 3.46 & 3.41   \\ 
  \hline
  \end{tabular}
  \caption{Comparison of values of $Z$ obtained from the five methods, for three sets of circumstances, described in the text.}
\end{table}

\subsubsection{Discussion}
Superficially, given that $\zbi=\zbf$ with Clopper-Pearson intervals is “Exact”, and the parametric bootstrap gives larger $Z$, the latter looks anti-conservative.
But recall that the so-called “exact” intervals overcover due to discreteness. Maybe $\zpb$ is actually better. We need a closer look.

\subsection{Coverage studies}

\begin{figure}[htbp]
\centering
\includegraphics*[width=2.7in]{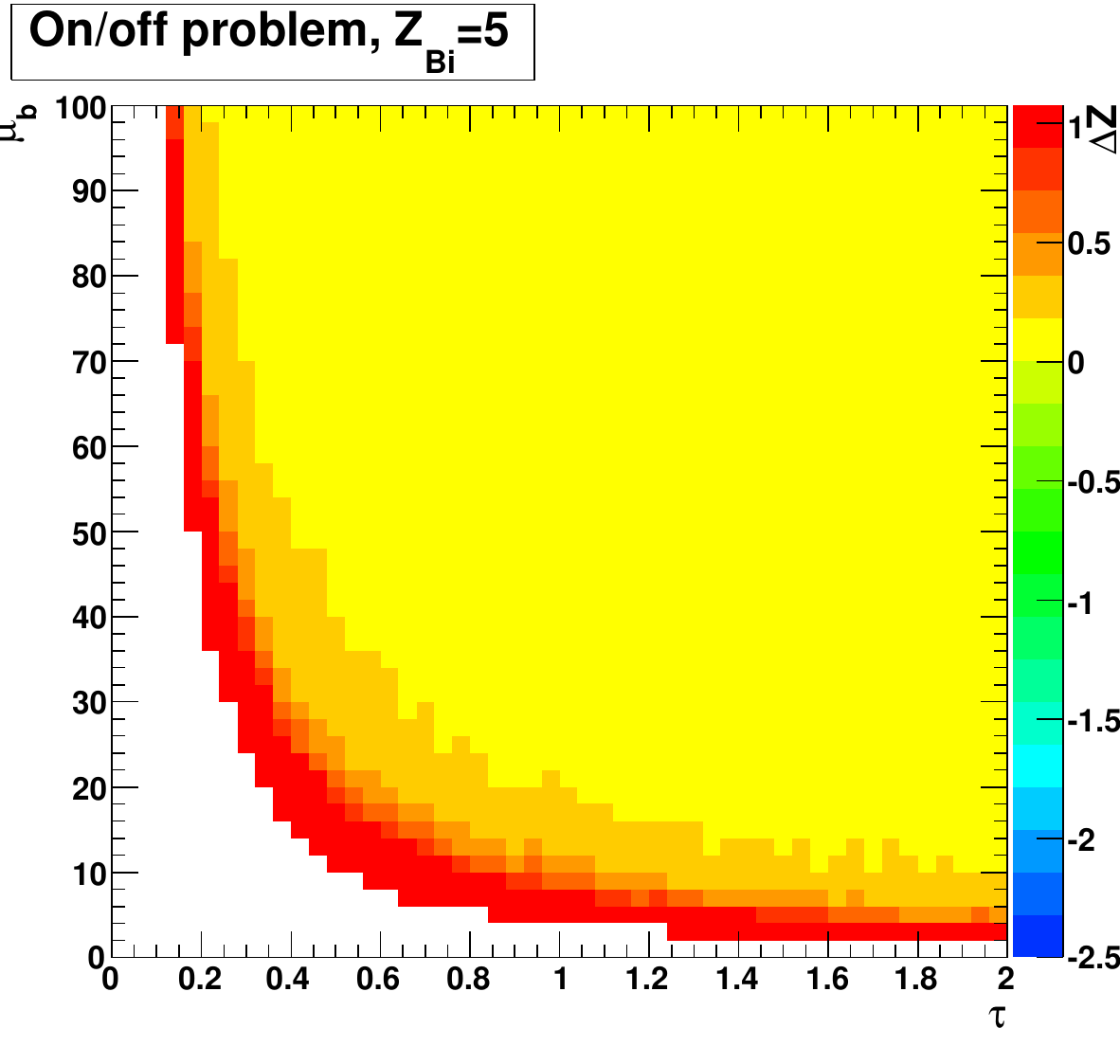}
\includegraphics*[width=2.7in]{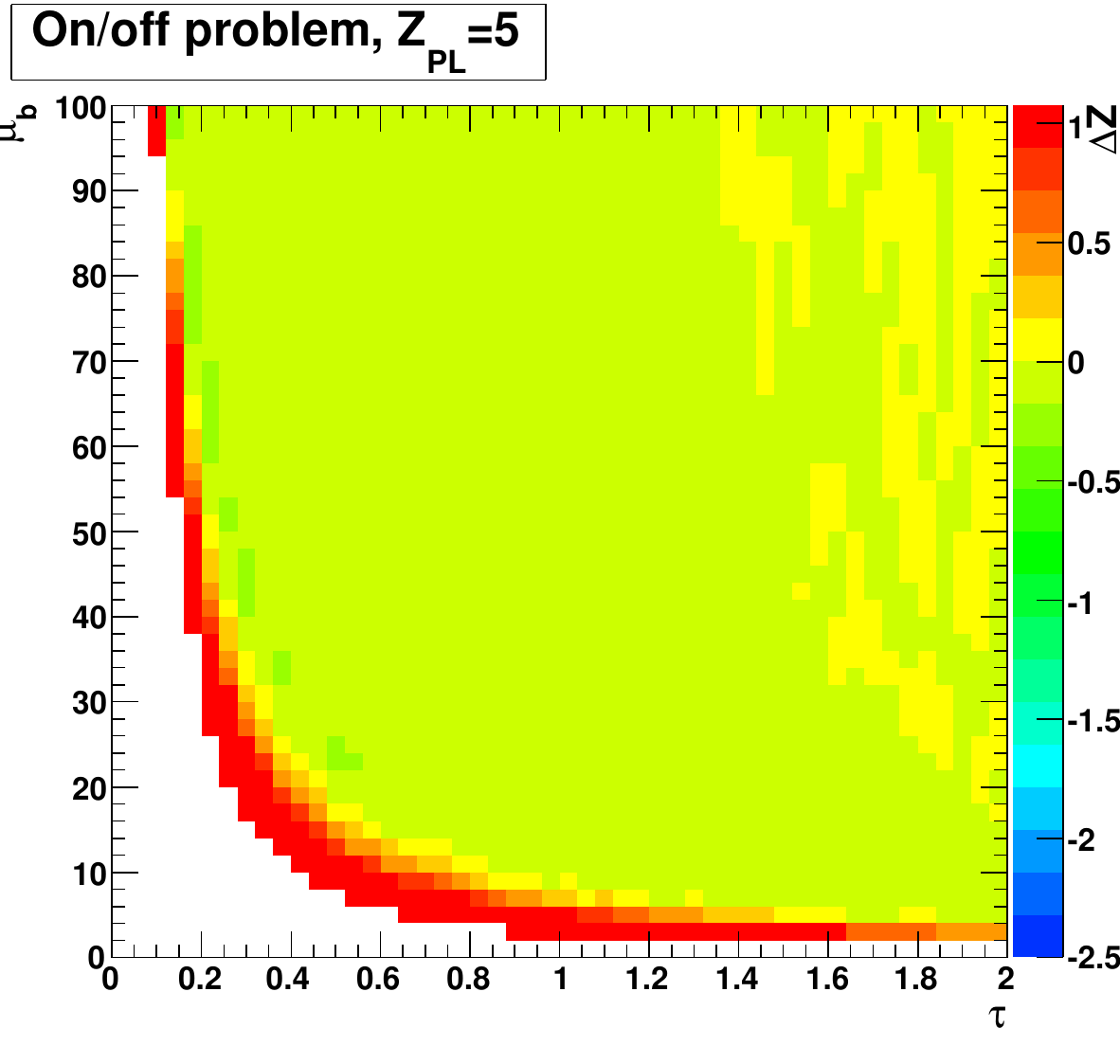}
\caption{For the on/off problem analyzed using (left) $\zbi$ recipe and
(right) the profile likelihood method, for
each fixed value of $\tau$ and $\mub$, the plot indicates
the calculated $\zdiff$ for the ensemble of experiments quoting
$\zclaim \ge 5$, i.e., a $p$-value of $2.87 \times 10^{-7}$ or
smaller. From Figs. 3 and 9 of \cite{CLT}.}
\label{proflik_lp_5}
\end{figure}

Following CLT, we consider a pair of true values of $(\mub , \tau)$, and for that pair,
consider all values of $(\non , \noff )$, and calculate both the probability of obtaining that data, and the computed $Z$ value for each recipe.  We consider some threshold value of claimed $\zclaim$, say 3 or 5, and compute the probability that
$Z\ge\zclaim$ according to a chosen recipe; this is the true Type I error
rate for that recipe and the significance level corresponding to that
value of $\zclaim$.  One can then convert this true Type I error
rate to a $Z$-value using the one-tailed formula and obtain what we call $\ztrue$.
This could be done by simulation, but we do direct calculations. Figure~\ref{proflik_lp_5} has ``heat maps'' from CLT showing $\zdiff$ for claims of 5$\sigma$ for the $\zbi$ and profile likelihood algorithms.

I calculated a couple of points using the parametric bootstrap $\zpb$:
For $\tau = 1.0$, $\mub = 10$, and $\zclaim = 3$, I found $\ztrue = 3.0$ (!)
For $\tau = 1.0$, $\mub = 20$, and $\zclaim = 5$, I found $\ztrue = 5.0$ (!)
Thus, at least for these two points, any anti-conservatism of the parametric bootstrap is exactly canceled out by the over-coverage due to discreteness.
That is, I cherry-picked my example problem to be one in which I knew that marginalizing the nuisance parameter (with judicious prior) yielded the standard frequentist $Z$, while doing parametric bootstrap as commonly done at the LHC would give higher $\zpb$. But in this example, the discrete nature of the data in the problem “saves” the coverage of $\zpb$ !

Note also the similarity of $\zpb$ with $\zmidp$. As mentioned, \cite{cht2010} found that Lancaster mid-P confidence intervals had excellent coverage in the unconditional ensemble, i.e., not conditioning on $\ntot$.

\subsection{\boldmath If $\tau$ is not known exactly}
An obvious next step is to relax the assumption that $\tau$ is known exactly, so that $\tau$ becomes a nuisance parameter. In analyses in HEP, $\tau$ is often estimated from the observed data in two so-called control regions (bins), which are thought to have the same ratio of background means $\tau$ as the original two bins.  The control regions are distinguished from the original two bins by using an additional statistic that is thought to be independent of the statistic used to define the original two bins. This is often referred to as the ABCD method, after the letters commonly designating the four bins.  I would strongly encourage coverage studies for the various methods applied to this case. One could also study the case where the estimate of $\tau$ is sampled from a lognormal distribution.

\section{Discussion at Banff BIRS 2023}
\label{sec:davison}

Among the many talks touching on the topic at Banff in \citeyear{davisonbanff2023}, statistician Anthony Davison’s was paired with mine, pointing to works including his Banff challenge paper with Sartori, and describing a parametric bootstrap, writing, 

``Lee and Young (2005, Statistics and Probability Letters) and DiCiccio and Young (2008, 2010, Biometrika) show that this parametric bootstrap procedure has third-order relative accuracy in linear exponential families, both conditionally and unconditionally, and is closely linked to objective Bayes approaches.''

\section{Final remarks}
\label{final}
Overall, I am encouraged by what I learned in the last year while preparing the talks and this writeup regarding our widespread use of the parametric bootstrap with the PLR test statistic in HEP.  Still, there is room to learn more (which I suspect that I must leave to the younger generations). First, how much is HEP losing by not studying more thoroughly the higher-order asymptotic theory in the test statistic?  Would that give more power, or at least reduce the need for parametric bootstrapping? Second, can one find relevant cases in HEP where marginalizing out-performs profiling?  Is there an example having continuous data but otherwise similar to the example in this note, so that one can study the performance without the confounding complication of discreteness?  Or, for those familiar with the ABCD method in HEP (where, as mentioned) $\tau$ is measured from contents of two more control regions), are there pitfalls for the parametric bootstrap in that context?
I would urge more exploration of these issues in real HEP analyses!

\section*{Acknowledgments}
Many thanks go to the multitudes of physicists and statisticians from whom I have learned, and particularly to Louis Lyons for bringing us together regularly at PhyStat meetings since the year 2000 (and for comments on earlier drafts). I also thank him, Sara Algeri, and Olaf Behnke for organizing this discussion. This work was partially supported by the U.S.\ Department of Energy
under Award Number {DE}--{SC}0009937.

\bibliographystyle{dcu}
 \bibliography{papers/Bob_refs}

\newpage
\addtocounter{section}{-6}
\begin{center}
{\bf \LARGE Part II: Statistician’s View}\\
\vspace{0.2cm}
by Larry Wasserman\\
\end{center}

\section{Introduction}

This paper arose
from my presentation
at the PhyStat meeting where
Bob Cousins and I were asked to
give our views on 
two methods for handling nuisance parameters
when conducting statistical inference.
This written account expands on my comments
and goes into a little more detail.

I'll follow Bob's notation where
$\mu$ is the (real-valued) parameter of interest
and $\beta$ is the vector of nuisance parameters.
In his paper,
Bob points out that
marginalising nuisance parameters
is rare at the LHC.
This is true in statistics as well.
There are a few cases where marginalising
is used, such as random effects models
where the number of parameters increases as the sample
size increases.
But even profile likelihood is not that common.
I would say that the most common method in statistics
is the Wald confidence interval
$\hat\mu \pm z_{\alpha/2} s_n$
where $\hat\mu$ is the maximum likelihood estimator,
$s_n$ is the standard error (usually obtained from the Fisher information matrix)
and $z_{\alpha/2}$ is the $\alpha/2$ upper quantile
of a standard Normal distribution.
(Note that for the Wald interval,
one simply inserts points estimates of the nuisance parameters
in the formula for $s_n$.)

Let
$Y_1,\ldots, Y_n$
be a random sample from
a density
$p_\theta$
which belongs to a model
$(p_\theta:\ \theta\in\Theta)$.
We decompose the parameter as
$\theta = (\mu,\beta)$
where $\mu\in\mathbb{R}$ is the parameter of interest and
$\beta\in\mathbb{R}^k$ is a 
vector of nuisance parameters.
We want to construct a confidence set
$C_n \equiv C_n(Y_1,\ldots, Y_n)$
such that
\begin{equation}\label{eq::coverage}
P_\theta(\mu\in C_n) \geq 1-\alpha\qquad \text{for\ all\ }\theta
\end{equation}
or, equivalently,
$$
\inf_\theta P_\theta(\mu\in C_n) \geq 1-\alpha.
$$
We may also want to test
$$
H_0: \mu = \mu_0 \qquad \text{versus} \qquad \mu\neq \mu_0
$$
however, for simplicity, I'll focus on confidence sets.
The present discussion
is on likelihood based methods.
In particular,
we may use the profile likelihood
$$
{\cal L}_{p}(\mu) = \sup_\beta {\cal L}(\mu,\beta)
$$
or the marginalised (or integrated) likelihood
$$
{\cal L}_{m}(\mu) = \int {\cal L}(\mu,\beta) \pi(\beta|\mu) d\beta
$$
where $\pi(\beta|\mu)$ is a prior for $\beta$.
In my view, we cannot say whether a likelihood is
good or bad 
without saying what we will do with the likelihood.
I assume that
we are using the likelihood 
to constructing
confidence intervals and tests.
Then judging whether a likelihood is
good or bad
requires assessing the quality of the confidence sets
obtained from the likelihood.
The usual central $1-\alpha$ confidence set based on ${\cal L}_p$ is
\begin{equation}\label{eq::prof}
C =\Bigl\{ \mu:\ {\cal L}_p(\mu) \geq {\cal L}_p(\hat\mu) e^{-c/2}\Bigr\}
= \Bigl\{ \mu:\ \ell_p(\mu) \geq \ell_p(\hat\mu) - \frac{c}{2}\Bigr\}
\end{equation}
where $\hat\theta = (\hat\mu,\hat\beta)$
is the maximum likelihood estimate,
$\ell_p = \log {\cal L}_p$ and
$c$ is the $1-\alpha$ quantile of a $\chi^2_1$
distribution.

\section{Profile or Marginalise?}

We want confidence intervals
with correct coverage
which means that
$P_\theta(\theta\in C)\geq 1-\alpha$ for all $\theta$.
In this case we say that
the confidence set is {\em valid}.
Among valid intervals,
we may want one that is shortest.
A valid
interval with shortest length
is {\em efficient}.
(But this is a scale dependent notion.)

If the sample size is large
and certain regularity conditions hold,
then the profile-based confidence set
in (\ref{eq::prof})
will be valid and efficient.
So in these well-behaved situations,
I don't see any reason
for using ${\cal L}_m$.
The main regularity conditions are:

(1) $p_\theta(y)$ is a smooth function of $\theta$.

(2) The range of the random variable $Y$ does not depend on $\theta$.

(3) The number of nuisance parameters does not increase
as the sample size increases.

In a very interesting paper that Bob mentions,
\cite{berger1999} 
provide examples
where ${\cal L}_p$ apparently does not behave well.
This is because,
in each case,
the sample size is small, or the number of nuisance parameters is increasing
or some other regularity conditions fail.
There may be a few cases where the marginalised
likelihood can fix these things
if we use a carefully chosen prior, but such cases
are rare.
We might improve the accuracy of the coverage
by using higher order asymptotics.
But this is rarely done in practice.
Such methods are not easy to apply
and they won't necessarily help
if there are violations of the regularity conditions.

Sometimes we can use specific tricks,
like conditioning on a well-chosen statistic
as Bob does in the Poisson ON-OFF example.
These methods are useful when
they are available but they tend to be very problem specific.
Bootstrapping and subsampling
are also useful in some cases.
A common misconception
is that the bootstrap is a finite sample method.
It is not.
These methods are asymptotic and they do
still rely on regularity conditions.

In the next two sections
I'll discuss two approaches
for getting
valid confidence sets
without relying on
asymptotic approximations or regularity conditions.

\section{Universal Inference}

Universal inference
\citep{wasserman2020universal}
is a method
for constructing
confidence sets
from the likelihood
that does not rely on
asymptotic approximations
or on regularity conditions.
The method,
works as follows.

\begin{enumerate}
\item Choose a large number $B$.
\item For $j=1,\ldots, B$:
\begin{enumerate}
\item Split the data into two disjoint groups
${\cal D}_0$ and ${\cal D}_1$.
\item Let $\hat\theta = (\hat\mu,\hat\beta)$
be any estimate of $\theta$ using ${\cal D}_1$.
\item Let
$T_j(\mu) = 
\frac{{\cal L}_0(\hat\mu,\hat\beta)}{\sup_{\beta}{\cal L}(\mu,\beta)}$
where
${\cal L}_0(\theta)$ is the likelihood 
constructed from ${\cal D}_0$ .
\end{enumerate}
\item Let $T(\mu) = B^{-1}\sum_{j=1}^B T_j(\mu)$.
\item Return $C_n = \{ \mu:\ T(\mu) \leq 1/\alpha\}$.
\end{enumerate}

This confidence set is valid,
that is,
$$
P_\theta(\mu\in C_n) \geq 1-\alpha\qquad\text{for\ all\ }\theta
$$
and this holds for any sample size.
It does not rely on large sample approximations
and it does not require
any regularity conditions.
The number of splits $B$ can be any number, even $B=1$ is allowed.
The advantage of using large $B$ is that it removes the
randomness from the procedure.
The disadvantage of 
Universal inference is that
one needs to compute the likelihood many times
which can be computationally expensive.

\section{Simulation Based Inference}

As the name implies,
simulation based inference (SBI)
uses simulations
to conduct inference.
It is usually used to deal with
intractable likelihoods
\citep{dalmasso2021likelihood, masserano2023simulator, al2024amortized,
cranmer2015, cranmer2020, thomas2022, xie2024,
masserano2022simulation, stanley2023}.
But it can be used as a way to get valid inference
for any model.
The idea is to first
get a confidence set for $\theta$
by inverting a test.
Let $\theta_*$ denote the true value of $\theta$.
For each $\theta$ we test the
hypothesis $H_0:\ \theta_* = \theta$.
This test is based on some test statistic
$T(\theta,Y_1,\ldots,Y_n)$.
The p-value for this test is
$$
p(\theta) = p(\mu,\beta)=
P_\theta\Bigl( T(\theta,Y_1^*,\ldots,Y_n^*) \geq T(\theta,Y_1,\ldots,Y_n)\Bigr)
$$
where
$Y_1^*,\ldots, Y_n^* \sim p_\theta$.
Then
$\tilde C = \{ \theta:\ p(\theta) \geq \alpha\}$
is an exact, $1-\alpha$ confidence set for $\theta$.
The confidence set for $\mu$
is the projection
$$
C = \Bigl\{ \mu:\ \theta = (\mu,\beta) \in \tilde C\ 
\text{for\ some\ }\beta\}\Bigr\}=
\Bigl\{\mu:\ \sup_\beta p(\mu,\beta) \geq \alpha\Bigr\}.
$$
Such projected confidence sets can be large,
but the resulting confidence sets have
correct coverage.
In SBI, we sample
a set of values
$\theta_1,\ldots, \theta_B$
from some distribution.
For each $\theta_j$
we draw a sample
$Y_1^*,\ldots, Y_n^* \sim p_{\theta_j}$.
Then we define
$$
I_j =I\Bigl( T(\theta_j,Y_1^*,\ldots,Y_n^*) \geq T(\theta_j,Y_1,\ldots,Y_n) \Bigr)
$$
which is 1 if
$T(\theta_j,Y_1^*,\ldots,Y_n^*) \geq T(\theta_j,Y_1,\ldots,Y_n)$
and 0 otherwise.
Note that
$$
p(\theta_j) = P_{\theta_j}(I_j = 1)=
\E[I_j | \theta_j].
$$
So $I_j$ is an unbiased estimate of
$p(\theta_j)$.
But we can't estimate $p(\theta_j)$
using a single observation $I_j$.
Instead, we
perform a nonparametric regression of
the $I_j$'s which allows us to use nearby values.
For example we can use the kernel regression estimator
$$
\hat p(\theta) =
\frac{\sum_j I_j K_h(\theta_j - \theta)}
{\sum_j K_h(\theta_j - \theta)}
$$
where $K_h$ is a kernel with bandwidth $h$
such as
$K_h(z) = e^{-||z||^2/(2h^2)}$.
Here are the steps in detail.

\begin{enumerate}
\item Choose any function
$T(\theta,Y_1,\ldots,Y_n)$.
\item Draw $\theta_1,\ldots,\theta_B$
from any distribution $h(\theta)$ with full support.
This can be the likelihood but it can be anything.
\item For $j=1,\ldots, B$:
draw $Y_1^*,\ldots,Y_n^*\sim p_{\theta_j}$ and compute
$T_j = T(\theta_j,Y_1^*,\ldots,Y_n^*)$.
\item Let
$I_j = I( T_j \geq T(\theta_j,Y_1,\ldots,Y_n))$
for $j=1,\ldots, B$.
Here,
$I_j=1$ if
$T_j \geq T(\theta_j,Y_1,\ldots,Y_n)$ and 
$I_j=0$ otherwise.
\item Estimate
$p(\theta_j) =\E[I_j | \theta_j]$ by regressing
$I_1,\ldots, I_B$ on
$\theta_1,\ldots,\theta_B$
to get $\hat p(\theta)=\hat p(\mu,\beta)$.
\item Return: $C = \{\mu\:\  \sup_\beta \hat p(\mu,\beta) \geq \alpha\}$. 
\end{enumerate}

We have that $P_\mu(\mu\in C_n)\to 1-\alpha$ as $B\to\infty$.
The test statistic
$T(\theta,Y_1,\ldots,Y_n)$
can be anything.
A natural choice is the
likelihood function
${\cal L}(\theta)$.
In fact, even if the likelihood is intractable
it too can be estimated using the simulation.
See
\citep{dalmasso2021likelihood, masserano2023simulator, al2024amortized,
cranmer2015, cranmer2020, thomas2022}
for details.
A related method is given in \cite{xie2022}.

\section{Optimality}

Suppose we have more than one valid
method.
How do we choose between them?
We could try to optimize some
notion of efficiency.
For example,
we could try to minimize
the expected length
of the confidence interval.
However, 
this measure is not transformation invariant
so one has to choose a scale.
The marginalised likelihood could potentially
lead to shorter intervals if the prior happens
to be concentrated near the true parameter value
but, in general, 
this would not be the case.
(\cite{hoff2022} discusses methods to include
prior information while maintaining frequentist coverage
guarantees.)
For testing, we usually
choose the test with the highest power.
But generally there is no
test with highest power at all
alternative values.
One has to choose a specific alternative or
one puts a prior on $\theta$ and maximizes
the average power.

Again, in the large sample,
low dimension regime
the choice of procedure might not have much
practical impact.
In small sample size situations,
careful simulation studies
are probably the best approach
to assessing the quality of competing
methods.

Of course, there may be other relevant criteria
such as simplicity and computational tractability.
These need to be factored in as well.

\section{Why Likelihood?}

There are cases where
neither the profile likelihood or the
marginalised likelihood should be used.
Here is a sample example.
Let
$c_1,\ldots, c_n$
be a set of fixed, unknown constants
such that
$0 \leq c_i \leq 1$.
Let
$\theta = n^{-1}\sum_i c_i$.
Let
$S_1,\ldots, S_n$
be $n$ independent flips of a coin
where
$P(S_i=1) = P(S_i=0) = 1/2$.
If $S_i=1$ you get to see $c_i$.
If $S_i=0$ then $c_i$ is unobserved.
Let
$$
\hat\theta = \frac{2}{n}\sum_{i=1}^n c_i S_i.
$$
Then
$\E[\hat\theta] = \theta$ and
a simple $1-\alpha$ confidence interval for $\theta$ is
$\hat\theta \pm z_{\alpha/2}/\sqrt{n}$.
The likelihood is
the probability of $n$ coin flips which is
$(1/2)^n$
which does not even depend on $\theta$.
In other words, the likelihood function is flat
and contains no information.
What's going here?
The variables
$S_1,\ldots, S_n$ are ancillary
which means their distribution
does not depend on the unknown parameter.
Ancillary statistics do not appear
in the likelihood.
But the estimator above does use the ancillaries
which shows that the ancillaries contain valuable information.
So likelihood-based inference
can fail
when there are useful ancillaries.
This issue comes up in real problems
such as analyzing randomized clinical trials
where the random assignment of subjects to treatment or
control is ancillary.

Another example
is Example 7 from
\cite{berger1999}.
Let $X_1,\ldots, X_n \sim N(\theta,\sigma^2)$
and suppose that
the parameter of interest is
$\mu = \theta/\sigma$.
\cite{berger1999}
shows that both profile and marginalised likelihood
functions have poor behavior.
But there is no need to use either.
We can define a confidence set
$C = \{ \theta/\sigma:\ (\theta,\sigma)\in \tilde C\}$
where $\tilde C$ is a joint confidence set for
$(\theta,\sigma)$.
This is a valid confidence set
which uses neither ${\cal L}_p$ or ${\cal L}_m$.

In some cases,
we might want to use semiparametric methods
instead of likelihood based methods.
Suppose, for example,
that the nuisance parameter is infinite dimensional.
For example, consider testing for the presence of a signal
in the presence of a background.
The data take the form
$$
Y_1,\ldots, Y_n \sim (1-\lambda) b(y) + \lambda s(y)
$$
where
$0 \leq \lambda \leq 1$,
$b$ is a known background distribution and
$s$ is the signal.
Suppose that $s$ is supported
on a signal region $\Omega$
but is otherwise not known.
We want to estimate $\lambda$ or test
$H_0: \lambda = 0$.
The unknown parameters are
$(\lambda,s)$.
Here, $s$ is any density such that
$\int_\Omega s(y)dy = 1$.
So $s$ is an infinite dimensional
nuisance parameter.
In our previous notation,
the parameter of interest $\beta$ is $\lambda$
and the nuisance parameter $\mu$ is $s$.

Models of this form
might be better handled using
semiparametric methods
\citep{tsiatis2006, kosorok2008}.
It is hard for me to judge whether such methods are useful
in physics but they may indeed be cases
where they are.
The main idea can be summarized as follows.
Suppose that $\mu$ is the parameter of interest
and $\beta$ is the (possibly infinite dimensional)
nuisance parameter.
Any well-behaved estimator
$\hat\mu$ will have
the property that
$\sqrt{n}(\hat\mu - \mu)$
will converge to a Normal
with mean 0 and some variance $\sigma^2$.
Generally,
$\sigma^2 \geq \E[ \varphi^2(Y)]$
for a function $\varphi$
which is known as the efficient
influence function.
We then try to find $\hat\mu$ so that
its corresponding $\sigma^2$ is equal to
$\E[ \varphi^2(Y)]$.
Such an estimator is said to be
semiparametric efficient.

In our signal detection example,
the efficient influence function is
$$
\varphi(y) = (1-\lambda)I_\Omega(y) -
\frac{P(\Omega)I_{\Omega^c}(y)}{\int_{\Omega^c} b(y) dy}
$$
where $P(\Omega) = P(Y\in\Omega)$
and the efficient estimator is simply
$$
\hat\lambda = 1-
\frac{n^{-1}\sum_i I_{\Omega^c}(Y_i)}{\int_{\Omega^c} b(y) dy}.
$$
The point is that we have taken care of an
infinite dimensional nuisance parameter
very easily
and we never mentioned
likelihood.

Another reason to consider
non-likelihood methods is robustness.
Both the profile likelihood and 
the marginalised likelihood could lead
to poor inferences
if the model is misspecified.
What should we do in such cases?
One possibility is to use
minimum Hellinger estimation.
In this case we
choose $\hat\theta$
to minimize
$$
h^2(\theta) = \int (\sqrt{p_\theta(y)} - \sqrt{\hat p(y)})^2 dy
$$
where $\hat p(y)$ is some non-parametric density estimator.
This estimator is asymptotically
equivalent to the maximum likelihood estimator
if the model is correct
but it is less sensitive
to model misspecification
\citep{beran1977, lindsay1994}.

Another recent non-likelihood method
is the HulC
\citep{hulc}.
We divide the data into
$B = \log(2/\alpha)/\log(2)$
disjoint groups.
Let $\hat\mu_1,\ldots,\hat\mu_B$
be estimates from the $B$ groups and let
$C =[\min_j \hat \mu_j,\max_j \hat\mu_j]$.
Define the median bias 
$$
B_n = \left(\frac{1}{2} - 
\min\Bigl\{ P(\hat\mu \geq \mu),\ P(\hat\mu \leq \mu)\Bigr\} \right).
$$
So $0 \leq B_n \leq 1/2$.
In the extreme
situation where
$\hat\mu$ is larger than the true value $\mu$
with probability 1,
we have $B_n = 1/2$.
Otherwise $B_n < 1/2$.
If $B_n\to 0$
as is typically the case,
we have
$P(\mu\in C_n) \to 1-\alpha$.
This is simple and does not require
profiling or marginalising a likelihood function.

\section{Conclusion}

The confidence intervals and tests
based on the profile likelihood
should have good coverage and reasonable size
as long as the sample size is large,
the number of nuisance parameters is not too large
and the usual regularity conditions hold.
If any of these conditions fail to hold,
it might be better to use
alternative methods such as
Universal inference,
simulation based inference,
semiparametric inference or the HulC.
The integrated likelihood may offer benefits
in some special cases
but I don't believe
it is useful in any generality.

I conclude with remarks about
power. 
Methods like Universal inference,
SBI and the HulC give satisfying coverage guarantees
but these come with some loss
of efficiency.
The confidence sets from Universal
inference typically shrink at the optimal
rate but will nonetheless be larger than optimal
confidence sets (if such optimal sets exist).
As a benchmark,
consider the confidence set
for
$Y_1,\ldots, Y_n \sim N(\mu,I)$
where $Y_i \in \mathbb{R}^d$.
Of course we don't need Universal inference here
but this case allows for precise results.
\cite{dunn2023} show that
the confidence set has radius of order $\sqrt{d/n}$
which is optimal but, compared to the best confidence set
the squared radius is about twice as large.
More precisely, the ratio squared radii is bounded by
$(4\log (1/\alpha)+4d)/(2\log (1/\alpha)+d -5/2)$.
Similarly, comparing the
HulC to the Wald interval
(when the latter is valid)
we find that the length is expanded by a factor of
$\sqrt{\log_2(\log_2(2/\alpha))}$ which is quite modest.
For SBI it is hard to say
much since that method works with any
statistic $T$ and the size of the set will depend
on the particular choice of $T$.
However, it is fair to say that
using generally applicable methods 
that provide correct coverage under weak conditions,
there will be some price to pay.
Evaluating these methods in some real physics problems
would be very valuable.
This would be a great project for
statisticians and physicists to collaborate on.

\section{Acknowledgments}

Thanks to Sara Algeri, Louis Lyons and
Olaf Behnke for organizing this discussion.
Also, I would like to thank
Louis Lyons and Bob Cousins for comments on an earlier
draft that led to improvements in the paper.

\bibliographystyle{dcu}
  \bibliography{papers/Larry_refs}

\end{document}